# Field induced superconductivity in a magnetically doped two-dimensional crystal

**Authors:** Adrian Llanos[1], Veronica Show[1], Reiley Dorrian[1], Joseph Falson[1,2]*

**Affiliations:**

[1] Department of Applied Physics and Materials Science, California Institute of Technology, Pasadena, California 91125, USA.

[2] Institute for Quantum Information and Matter, California Institute of Technology, Pasadena, California 91125, USA.

*Corresponding author. Email: falson@caltech.edu

**Magnetic field induced superconductivity is a rare property in nature due to the sensitivity of spin-singlet Cooper pairing to time-reversal symmetry breaking perturbations. However, in rare cases, an interplay between magnetic fields and ions can be engineered to bring about superconductivity at finite fields. Here we use ultra-thin $LaSb_2$ doped with dilute Ce paramagnetic impurities to demonstrate a magnetic field-induced superconducting dome in a two-dimensional crystal. The reduced dimensionality of the structure enables the use of an in-plane magnetic field to dynamically suppress spin fluctuations on the Ce-site, which leads to an anomalous enhancement of the critical temperature with increasing field. By modelling the spin scattering dynamics across the experimental parameter space, we reveal insight into the complex nature of paramagnetic impurities in magnetic fields at low temperature, and how their manipulation can result in the ability to tune between competing magnetic pair-breaking regimes. Realizing this physics in a two-dimensional crystalline setting invites the application of similar approaches to unconventional forms of superconductivity while also highlighting new experimental standards which should be employed when studying ultra-thin materials in general.**

The centrality of symmetry in the physics of superconductivity was realized soon after the development of the Bardeen–Cooper–Schrieffer (BCS) theory. The combination of time-reversal and inversion symmetry, responsible for the two-fold spin-degeneracy of the conduction electrons, is essential for preserving *s*-wave pairing in conventional superconductors, even in the presence of strong disorder (*1*). Accordingly, superconductors with broken symmetries are prime platforms for probing unconventional superconducting behavior. Both the breaking of inversion symmetry in the presence of strong spin-orbit coupling as well as time-reversal symmetry breaking leading to spin polarization have been used as pathways to conceive microscopic mechanisms such as Ising (*2*), spin-triplet (*3*), or finite momentum pairing (*4,5*). Experimental canvassing of an array of materials has also tied the importance of symmetry to exotic pairing through the observation of extremely high upper critical fields (*6-10*), large critical field anisotropies (*11-13*), reentrant (*14-17*), and even field-induced phases in magnetically ordered crystals (*18-21*). These phases are often probed by utilizing two-dimensional (2D) crystals, as the influence of anisotropies and crystalline symmetries are amplified in reduced dimensions while the suppression of the superconducting state in a magnetic field applied parallel to the film is weakened. And while changing layer number and chemical composition have been extensively employed for the tuning of superconducting properties, in this work, we study the impact of competing time reversal symmetry breaking perturbations by way of precision doping dilute paramagnetic impurities into a 2D crystalline superconductor.

The platform for our study is the recently discovered 2D superconductor, monoclinic $LaSb_2$ (*22*) which here we intentionally dope with Ce. Initially addressed by Abrikosov and Gor'kov (AG) (*23*), magnetic ions are known to break time-reversal symmetry and destroy superconductivity through exchange scattering of the conduction electrons (*24,25*). Of interest here is the rich physics of this scattering process in the presence of an applied *H*-field at very low temperatures. While an external *H*-field suppresses superconductivity by coupling to the orbital (known as the orbital effect) and also to spin degrees of freedom (paramagnetic effect), the field will also polarize the paramagnetic impurities. The result of this is twofold. First, polarization of the ions generates an internal exchange field (*26-28*) which can compensate or augment the applied *H*-field. The second effect is on scattering of carriers from fluctuating ions. While early theoretical studies suggested that exchange scattering rate could be suppressed through polarization in a *H*-field (*29-31*), modern extensions of the problem (*32,33*) have predicted exotic scenarios of *field-induced* superconductivity emerging when exchange scattering is suppressed in disordered isotropic metals with sufficiently weak orbital and paramagnetic depairing effects. This behavior is yet to be observed to date in a crystalline 2D material, with the only indications resolved in ultra-narrow wires (*34*) and doped amorphous Pb films (*35*). Thus, it has remained unclear whether crystalline systems which host longer mean-free paths and reduced spin orbit scattering (SOS) can be suitable hosts for such physics. Here we use 2D crystals of $LaSb_2$ doped with a critical concentration of Ce to chart a field-induced superconducting dome with a fully-developed zero-resistance state emerging at finite parallel fields. We illustrate the ability to dynamically control the spin exchange scattering rate as a key ingredient in enabling this physics.

The unit cell of $LaSb_2$ is shown in Fig.1**a** and is characterized by a monoclinic shear between two quintuple layer (QL) building blocks of the structure. The normalized resistance $R/R_0$ data plotted in Fig.1**b** illustrates that the thin film samples superconduct down to a thickness (*t*) of 4.4 nm, which corresponds to five QL. A notable outcome of the monoclinic structure is the enhanced critical temperature $T_c$ compared to the bulk (*36*), which we attribute to an increased Josephson coupling between the quasi-van der Waals QL in the structure due to the ~5% compressed *c*-axis compared to that found in bulk crystals. The 2D nature of superconductivity in a 4.4 nm film is demonstrated by mapping $H_{c2}$ as a function of angle $\theta$ relative to the in-plane direction. The data shown in Fig.1**c** shows a sharp cusp when the field is applied parallel to the sample ($\theta=0°$) in agreement with the behavior expected of a 2D superconductor according to a 2D Tinkham analysis (Eq. S1), as plotted by the dotted red line *(37)*. The contrasting 3D anisotropic Ginzburg-Landau analysis (Eq. S2) is presented as a blue dotted line. Resilience to $H^{\parallel}$ is demonstrated by $H_{c2}$ exceeding the Chandrasekhar-Clogston or Pauli limit $\mu_0 H_{Pauli}=1.86T_c$, indicated by the horizontal line in Fig.1c under the assumption of the *g*-factor equaling 2. Several examples of materials exceeding this limit have been found with mechanisms varying from strong SOS to Ising superconductivity (*38*). An analysis of the out-of-plane critical fields (Fig.S3) allows us to estimate the Ginzburg-Landau coherence length $\xi$ of samples which exceeds the film thickness shown in Fig.1d. Despite the increase in observed $T_c$, the coherence length is found to increase with sample thickness, which we attribute to the decrease in normal state sheet resistance. This behavior is expected for dirty-limit superconductors as the mean free path length increases (*37*). Taken together, this data presents LaSb$_2$ as a new example of a 2D superconductor.

We now take advantage of the solubility of Ce onto the La site in the crystal to probe the influence of magnetic perturbations on the 2D superconducting state. In our experiments, we control the La/Ce flux ratio during growth by modifying the Ce effusion cell temperature in order to achieve a range of Ce concentrations (*c*). X-ray diffraction confirms that all samples remain highly crystalline (Fig.S1). Two sets of samples exhibit similar physics and are shown in the supplementary information (SI). Since reliable quantification of *c* is generally challenging in thin samples, here we have chosen to focus on one set of samples from a single growth run. A gradual suppression of $T_c$ with increased Ce is resolved, as shown in Fig.2a and summarized in Fig.2b. Notably, by $c \approx 0.019\%$ (in atomic percent) we are able to suppress superconductivity, leaving only an incipient downturn at very low temperatures. With further doping to $c \approx 0.025\%$, we observe an upturn in resistance as $T \rightarrow 0$ indicative of a transition to a weakly localized metal.

The reduction in $T_c$ at zero field by magnetic dopants is given by the AG formula (*23*):

$$ln\left(\frac{T_{c0}}{T_c}\right) = \psi\left(\frac{1}{2} + \frac{\hbar v_s}{2\pi k_B T_c}\right) - \psi\left(\frac{1}{2}\right) \quad (1)$$

Where $T_{c0}$ is the impurity-free $T_c$, and $v_S$ is the magnetic-impurity exchange scattering rate which is given by:

$$v_s = 2\pi N_F n_s J_{ex}^2 J(J+1) \quad (2)$$

and depends on the impurity concentration as well as the exchange coupling constant $J_{ex}$, $n_s$ the impurity concentration, the density of states $N_F$ of the metal host and in the present case, where SOC is strong, the impurity total angular momentum $J$. $\psi$ is the digamma function, $\hbar$ is the reduced Planck constant, and $k_B$ is the Boltzmann constant. These formulae imply $v_S$ determines $T_c$ for a doped sample and is the most physically relevant parameter. Henceforth, we will refer to samples by this value.

In Figure 2c we plot the $T_c$, defined as the half-point in normal resistance, in the ($H^{\parallel}$,$T$)-parameter space for the 5 QL set of samples labelled by the respective value for $v_s$. For the lightly doped samples (small $v_S$), a suppression of $T_c$ with increasing $H^{\parallel}$ is observed and can be fit by an AG-like formula with an appropriate $H$-dependent pair-breaking rate from the orbital and paramagnetic effects due to the external field (*38,39*). By $v_s$=0.60$T_{c0}$, the response to the field is qualitatively different, with a modest increase in $T_c$ when increasing $H^{\parallel}$, pointing to field-enhanced superconductivity. The field response of the 0.89$T_c$ sample is even more striking, as it is not superconducting at zero field yet develops into a full zero-resistance state upon application of $H^{\parallel}$ with a maximum $T_c \approx$ 120 mK at $H^{\parallel} \approx$0.6 T. This is our demonstration of field-induced superconductivity driven by a parallel magnetic field for samples doped with critical concentration of magnetic impurities. Fits to all these data, indicated as solid lines for each data set, were obtained using the Kharitonov-Feigelman (KF) theory (*32*) which will be discussed in detail later. For the $v_S$=0.89$T_c$ sample, the resistance as a function of temperature at a constant in-plane field is displayed in Fig.2**d** and as a function of the magnetic field at constant temperatures in Fig.2**e**. Additional datasets including a second sample series are presented in the SI (Fig.S7,S8).

We now use an out-of-plane magnetic field $H^{\perp}$ in addition to $H^{\parallel}$ to study the $v_S$= 0.89$T_c$ sample in detail. Applying $H^{\perp}$ rapidly suppresses superconductivity through the orbital effect. Figure 3**a** displays $R(T)$ with finite $H^{\perp}$ while $\mu_0 H^{\parallel}$ = 0.6 T, corresponding to the "nose" in Fig.2**c**. A value for $\mu_0 H_{c2}^{\perp}$~7mT is seen for the zero resistance state, which evolves into a metallic state characterized by finite resistance in the limit of $T \rightarrow 0$ (*40*). By plotting $H_{c2}^{\perp}(T)$ at discrete values of $H^{\parallel}$ as shown in Fig.3**b**, fits to a single band Werthamer, Helfand, and Hohenberg (WHH) (*41*) model provide an estimate for $H_{c2}^{\perp}$ ($T$=0), allowing us to obtain $\xi$ as a function of $H^{\parallel}$. We find that values of $H^{\parallel}$ on opposite sides of the dome with similar $T_c$ do not yield the same $H_{c2}^{\perp}(T)$, as shown in Fig.3**c**. Instead, the values obtained on the high-field side are systematically lower. A similar asymmetry is apparent in the values of the London penetration depth $\lambda$, extracted using the model (Eq. S6) for the collective creep of vortices (*42, 43*).

We now illustrate how these effects can be understood to result from the dynamical response of the magnetic impurities to an applied field. Extending the work of AG, Maki showed that all pair-breaking mechanisms are "equivalent" in their effect on $T_c$, i.e., they can all be shown to follow an AG type equation with a generalized $v_s$ which depends on the specific conditions (*44*). This line of reasoning is adopted by the KF framework (*32*) which we have used to model our data. While assuming isotropic, *s*-wave pairing, the KF theory extends AG to include finite polarizability of impurity spins and their effects on both the exchange scattering (within the Born approximation) as well as the generated internal exchange field when polarized. The polarization of impurities due to the applied field leads to a decrease in the exchange scattering pair-breaking rate $\Gamma$ while simultaneously degrading superconductivity through the orbital and paramagnetic effects. This competition is captured by calculating a modified Cooper pair Green function (Eq. S11) with contributions to the lifetime from each of these effects, which in turn determines $T_c$. A fit to the 0.89$T_c$ sample data using antiferromagnetic exchange coupling ($J_{ex}$>0) is shown in Fig.4**a,** with fits to extended data sets given in the SI. Given the 3+ valence of La in this structure, we assign a similar valence for the Ce ions, leaving a *4f¹* configuration for the remaining valence electron and $J = 5/2$. For simplicity, we have ignored both crystal field splitting of the localized *4f* electrons as well as Kondo screening of the impurity by the conduction electrons. We find the data can be well described by this model with physically reasonable values for the fit parameters. We present additional details of the fitting procedure in the SI.

It is $\Gamma$ that we now study in more detail, as its behavior when tuning temperature and applied field reproduces the features of our data. The exchange scattering rate $\Gamma$ is independent of the sign of $J_{ex}$, but depends strongly on the degree of spin polarization, that is, on $T$ and $H$ through the Brillouin function. Using the KF theory, we plot the calculated $\Gamma/v_s$ ratio as a color contour in Fig.4**a**, with line cuts taken at $\mu_0 H^{\parallel}=0.1$ and 1.1 T shown in Fig.4**b**. When $H^{\parallel}/T$ is small, the external field is insufficient to polarize the magnetic impurity spins. This corresponds to the upper left corner of Fig.4**a**, where $\Gamma$ equals $v_s$ as expected. When $H^{\parallel}/T$ is large, however, spins are readily polarized by the large field. In this limit, Zeeman splitting of the impurity level suppresses the spin-flip contribution to the exchange scattering, which in turn reduces the full exchange scattering rate $\Gamma$ from $v_s$ to $v_s J/(J+1)$. This is the lower right of Fig.4**a**. This mechanism is responsible for the observed enhancement of $T_c$ in an applied parallel field. However as the field increases, $\Gamma$ is reduced to a minimum, the orbital and paramagnetic contributions to pair breaking come to dominate, and $T_c$ is again suppressed.

We extend this line of reasoning to understand the anomalous perpendicular field dependence. At low $H^{\parallel}/T$, $\Gamma$ continues to fall below $T_c$, with the resulting suppression of spin-flip scattering enhancing superconductivity, and therefore the critical field, in the limit of $T \rightarrow 0$ (45). In contrast, at large $H^{\parallel}/T$ the external field has already nearly fully polarized spins at $T_c$ leaving $\Gamma$ with only a weak temperature dependence. Despite yielding similar $T_c$, values of applied parallel field on opposite sides of the dome give rise to differing dynamical responses of *competing* pair-breaking mechanisms to temperature and field and result in the observed asymmetric responses of $H_{c2}^{\perp}$, $\xi$ and $\lambda$, obtained in the limit of $T \rightarrow 0$.

We have demonstrated the possibility of realizing a full field-induced superconducting phase in the ($H^{\parallel}$,$T$)-parameter space in a chemically versatile 2D crystal. Modelling of the data provides insight into competing depairing mechanisms and their dynamic nature. Our analysis serves as a starting point for a renewed investigation of this physics, and for future studies which tackle the influence of Fermi surface anisotropies, broken symmetries and correlation effects (e.g. Kondo screening, Shiba states). Importantly, our data demonstrate that even in the absence of exotic scenarios such as Ising, spin polarized, or finite momentum pairing as leading explanations, a similarly rich variety of tunable behaviors can manifest when competing pair-breaking mechanisms are present. This has implications for the interpretation of qualitatively similar anomalous behaviors elsewhere, especially when there is a possibility of magnetic ion incorporation into the structure which may even be unintentionally incorporated from impure source materials. The work also highlights the necessity of studying ultra-thin 2D materials in a finite in-plane field to establish the presence or absence of superconductivity. It is probable that a multitude of materials show similar physics and have evaded detection.

## Methods

### Film growth

Samples were prepared using a molecular beam epitaxy apparatus at a base pressure of $10^{-10}$ mbar. Films were grown using nominally the same recipe as reported in *(22)* on thermally prepared MgO (001) substrates. Samples were prepared in two synthesis runs, punctuated by machine maintenance and upgrades to chamber capabilities. The samples are summarized in Table S1. All growths were performed with similar growth conditions. The growth begins with a low temperature buffer layer with La and Sb codeposited at beam flux pressures of $6.5 \times 10^{-9}$ mbar and $1.5 \times 10^{-7}$ mbar, at growth temperature 315°C. The buffer layer is annealed at 615°C for 5 minutes before being cooled to 520°C, where the rest of the film is grown with La and Sb codepositing at beam flux pressures of $1 \times 10^{-8}$ mbar and $1.5 \times 10^{-7}$ mbar. The growth rate was estimated using Laue fringes observed in lower thickness films and was found to be 0.048 Å/s, as shown in Fig.S1. To reduce degradation in air, all films were capped with amorphous Ge *in-situ* at room temperature. Cerium concentration was estimated using secondary ion mass spectroscopy by establishing a correlation between effusion cell temperature and doping concentration in densely doped samples, and extrapolating this to the dilute limit (see SI for details).

### Transport measurements

Transport measurements were obtained in a dilution refrigerator equipped with a 2-axis vector magnet (9-3T). Samples are in vacuum and are probed using thermally anchored thermocoax lines with additional low-temperature high frequency filters with cut off frequency of 20 kHz. The sample temperature is monitored using both a resistive ruthenium oxide and paramagnetic salt thermometer. The base temperature of the mixing chamber is around $T = 7.5$ mK, with the sample base temperature around 15 mK. Transport data was obtained on macroscopic samples approximately 5 mm$^2$ which are directly bonded using Al wire, with a linear 4 probe contact arrangement. Alternating current excitations were 10 nA were used below 150mK, although 100 nA was used at higher temperatures (>300mK) in cases where a reduction in noise was required, and voltage drops were measured using lock-in amplifiers operating between 10 and 20 Hz.

**Acknowledgments:** We appreciate discussions with Mikhail Feigelman. We thank Dr Yunbin Guan and the Caltech Microanalysis Center for secondary ion mass spectroscopy measurements and the Beckman Institute for their support of the X-Ray Crystallography Facility at Caltech.

**Funding:**

Air Force Office of Scientific Research Grant number FA9550-22-1-0463 (JF)

Gordon and Betty Moore Foundation's EPiQS Initiative Grant number GBMF10638 (JF)

Institute for Quantum Information and Matter, an NSF Physics Frontiers Center, Grant PHY-2317110 (JF).

**Contributions**

A.L. and J.F. conceived the project. A.L., V.S. and R.D. synthesized and characterized films. A.L. and J.F. performed the low temperature measurements. A.L. performed fits and analyzed the data. A.L. and J.F. prepared figures. A.L. and J.F. wrote the manuscript with assistance from V.S. and R.D.

**Competing interests:** Authors declare that they have no competing interests.

**Data and materials availability:** Data and code is available at the Caltech DATA server: https://data.caltech.edu/

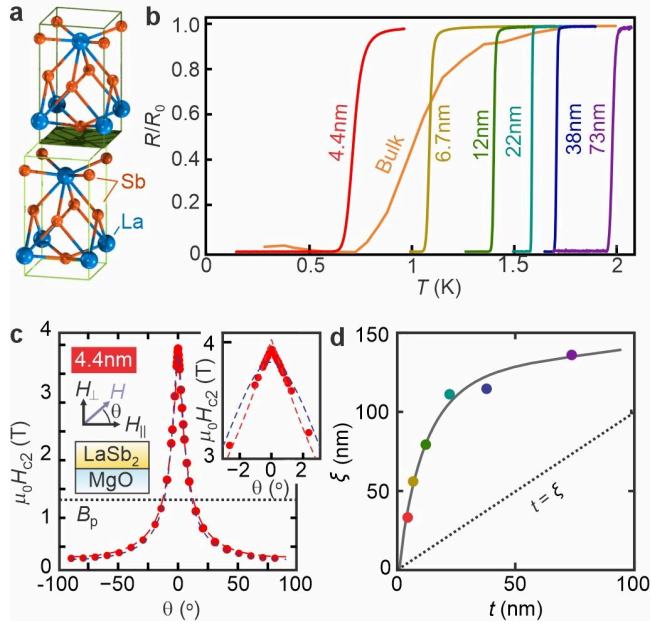

**Figure 1. 2D superconductivity in LaSb$_2$ thin films. a** Illustration of the unit cell of LaSb$_2$ composed of two sheared quintuple layers. **b** Thickness dependent resistive transitions of the superconducting state. Bulk data is adapted from Ref. *(36)*. **c** Tinkham analysis of a thin film in an in-plane magnetic field. The two dotted lines correspond to a 2D (red) and 3D (blue) GL function. **d** Coherence length $\xi$ obtained as a function of film thickness.

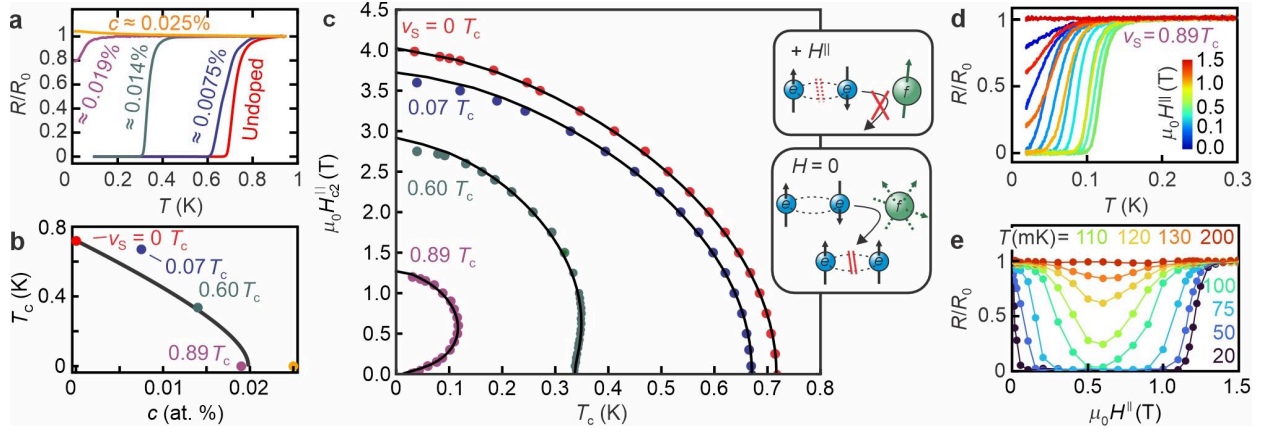

**Figure 2. Field-induced superconductivity in Ce doped LaSb$_2$.** **a** $R/R_0$ as a function of films with $t$ = 4.4nm of different cerium concentrations $c$. **b** $T_c$ as a function of $c$ with the spin exchange rate $v_s$ obtained from the KF formula noted for each sample. **c** In-plane critical field $H_{c2}^{\parallel}$ as a function of $T$ for the cerium doped samples. Solid lines correspond to fits according to the KF model. The $H = 0$ (lower) schematic shows a singlet Cooper pair interacting with a randomly fluctuating $f$-electron leading to a scattering process which flips spin and breaks the Cooper pair. The $+H^{\parallel}$ (upper) schematic shows that application of a field polarizes the $f$-electron, suppressing the scattering channel, but leading to pair breaking between paired electrons due to the applied field. **d** $R/R_0(T)$ of sample $v_s$=0.89$T_c$ at various $H^{\parallel}$ and **e** as a function of $H^{\parallel}$ at various $T$.

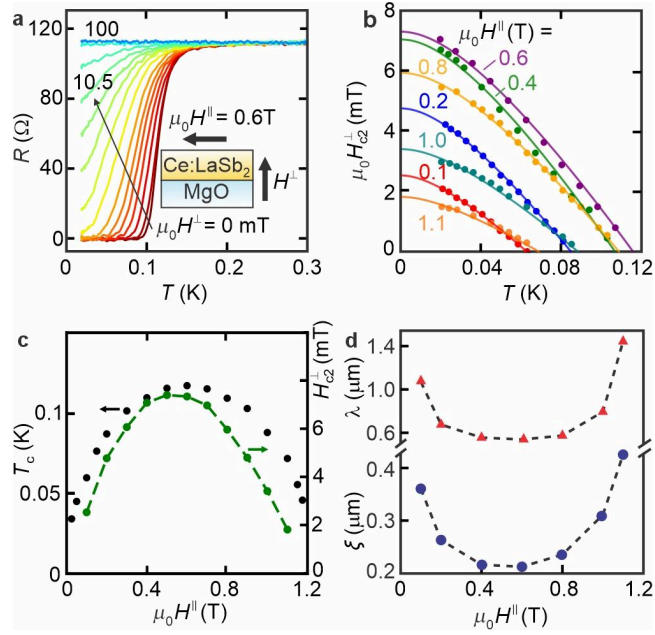

**Figure 3. Perpendicular Critical Field Behavior for sample $v_s=0.89T_c$.** **a** $R(T)$ taken in a fixed $H^{\parallel} = 0.6$ T in a set of $H^{\perp}$ from 0T to 0.1 T. Curves taken between 0 and 10.5mT were taken at 1-2mT intervals. **b** $H^{\perp}_{c2}$ as a function of temperature taken at different $H^{\parallel}$. Solid lines are fits to the WHH model (*41*). **c** $T_c$ and $H^{\perp}_{c2}$ plotted as a function of $H^{\parallel}$. **d** Summary of $\xi$ (circles) and $\lambda$ (triangles) as a function of $H^{\parallel}$.

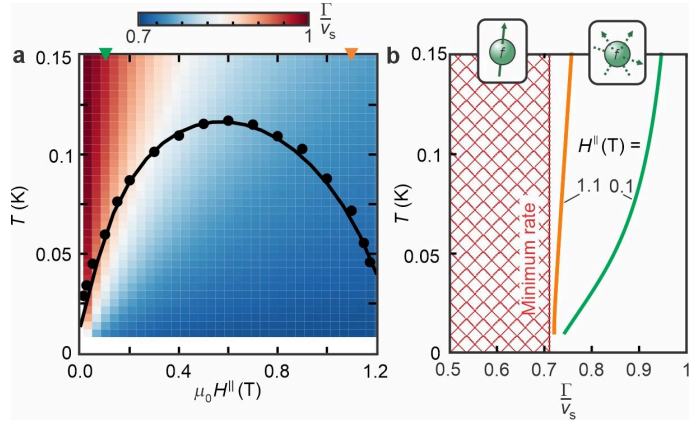

**Figure 4. Summary of data and fitting parameters for sample $v_s=0.89T_c$. a** $T_c$ (black circles) and KF fit (black solidline) as a function of $H^\parallel$. The color contour presents $\Gamma/v_s$ plotted in the $(H^\parallel,T)$-parameter space. **b** Line cuts of $\Gamma/v_s$ at $H^\parallel=0.1$ (green line) and 1.1T (orange line) as a function of $T$.

# Supplementary Information for

# Field induced superconductivity in a magnetically doped two-dimensional crystal

**Authors:** Adrian Llanos[1], Veronica Show[1], Reiley Dorrian[1], Joseph Falson[1,2]*

**Affiliations:**

[1]Department of Applied Physics and Materials Science, California Institute of Technology, Pasadena, California 91125, USA.

[2] Institute for Quantum Information and Matter, California Institute of Technology, Pasadena, California 91125, USA.

*Corresponding author. Email: falson@caltech.edu

**The PDF file includes:**

>   Summary of Samples
>   Field Alignment Procedure
>   Cerium Concentration Analysis
>   Supplementary Text
>   Figs. S1 to S8
>   Tables S1 to S4

**Summary of samples**

**Table S1. Summary of synthesis equipment and source materials for series 1 & 2.**

| Samples | Source cells | Source Material |
|---|---|---|
| *Series 1:*<br>$T_{c0}$ = 717 mk<br>$v_s$ = 0, 0.07, 0.6, 0.89 | La: High T effusion cell<br>Ce: High T effusion cell<br>Sb: Hot-lip effusion cell | La: 4N Ames laboratory (first ingot)<br>Ce: 3N ACI alloys<br>Sb: 6N Heeger Materials |
| *Series 2:*<br>$T_{c0}$ = 838 mK<br>$v_s$ = 0, 0.56, 0.75, 0.84, 0.89 | La: High T effusion cell<br>Ce: High T effusion cell<br>Sb: Thermal cracker | La: 4N Ames laboratory (second ingot)<br>Ce: 3N ACI alloys<br>Sb: 6N Thermofisher Scientific |

The films are twinned in-plane due to the square symmetry of the template MgO (001) substrate, and therefore we are unable to detect differences between *a*- and *b*-axis transport features. X-ray diffraction on two films are shown in Fig. S1. Reflected high energy electron diffraction and atomic force microscopy images from a 4.4nm film are shown in Fig.S2. The diffraction pattern is streaky and has high-contrast indicating good surface crystallinity. AFM was used to assess the quality of the surface where it can be seen that the surface consists of large, flat regions punctuated by deeper cavities.

**Field alignment procedure**

Samples were mounted vertically with the sample normal pointing parallel to the 3T, Y-magnet direction. For the measurements of the temperature dependence of resistance in a fixed parallel field (Fig. 2 of the main text), the field was first aligned by sweeping the vector angle at a temperature close to $T_c$. The condition of optimal alignment was found by locating the angle which gave the minimum resistance. Sample tilt in the Y-Z plane was found to be ~1°. The third axis could not be aligned, but deviation in that direction was estimated to be ~≲5°.

For the measurements of the perpendicular critical field, the 9T Z-magnet was kept at a fixed value (corresponding to an in-plane field) while the Y-magnet was swept at 2mT/min at fixed temperature (corresponding to an out-of-plane field). Due to the combination of trapped fluxes and sample misalignment, sweeps were found to be symmetric about non-zero values (≲10mT) of the Y-field. Data were therefore corrected with this value set to be true zero. Given the small critical field values of <10mT, parasitic contributions from the Y-field in the parallel direction would amount to <1mT corresponding to < 1% uncertainty in the parallel field magnitude. For consistency, temperature dependent sweeps in a fixed parallel and perpendicular fields were also performed by fixing both the Z magnitude and varying the Y magnitude in ~0.1–1mT steps for each sweep. The condition of zero perpendicular field was given by the Y field value that gave the maximum $T_c$. Temperature dependence of the critical field obtained by sweeping *H* at fixed *T* and by sweeping *T* with fixed *H* were in agreement.

**Cerium concentration analysis**

The cerium concentration was estimated using secondary ion mass spectroscopy (SIMS) at the Caltech Microanalysis Center. A -12.5 kV $^{16}O^-$ primary ion beam of ~10 nA (~20 um in diameter) was used to sputter the samples in rastering mode (100x100 μm) to produce secondary ions. A field aperture of 300 μm in diameter was used to enable only the ions from the central ~40 μm of the craters to be transmitted for detection. Possible edge effect was further eliminated with a 63% (in area) electronic gating. Secondary ions ($^{139}La^+$ and $^{140}Ce^+$) of 8.5 keV were collected with an electron multiplier (EM) in the peak-jumping mode. Each measurement consisted of about 130 cycles of data collection, reaching the steady state of the secondary ion production. The counting time of each mass was 1 sec per cycle. The energy bandwidth for the secondary ions was set at ~45 eV. Data were corrected for EM background and deadtime. We found that the technique was unable to reasonably quantify the Ce:La ratio directly for the ultra-dilute films studied in the text ($c \ll 1\%$). In addition, the Ce beam flux during the growth of these films was too small to measure using a beam flux monitor. The Ce concentrations were therefore determined using SIMS results of more concentrated films ($c$ = 1~25%). This data was organized on an Arrhenius-type plot to correlate the atomic concentration of Ce with the temperature of the Ce effusion cell (Fig.S3). Extrapolating this fit to cell temperatures in the range 1100-1150 ºC provides the estimated Ce concentrations used throughout the text which we put in the range of $c \leq 0.05$ atomic %. We note, however, that while the overall trend of increasing Ce results in suppression of $T_c$, there is some scatter in our estimate of concentration. This could be due to a couple of effects. Firstly, as we can not measure flux directly at such low levels, it is possible that there are growth-to-growth fluctuations in the Ce flux. Also, these growths have occurred over the space of approximately six months where the La and Ce cell was used extensively and is kept as a liquid inside the MBE crucible. This can result in a natural distillation of source materials leading to higher purities over time. Thus, while we can be confident in the order of magnitude concentration, it is challenging to make highly accurate quantitative correlations between $T_c$ and $H_c(T)$ and $c$ in our current experimental approach.

**Supplementary Text**

**Analysis of 2D Superconductivity**

2-D superconductivity was verified by fitting the angle-dependent critical field to the 2D Tinkham formula:

$$\frac{H_{c2}(\theta)|\sin(\theta)|}{H_{c2}^{\perp}} + \left(\frac{H_{c2}(\theta)\cos(\theta)}{H_{c2}^{\parallel}}\right)^2 = 1 \quad \text{(Eq.S1)}$$

and to the 3D anisotropic Ginzburg-Landau formula

$$\left(\frac{H_{c2}(\theta)\sin(\theta)}{H_{c2}^{\perp}}\right)^2 + \left(\frac{H_{c2}(\theta)\cos(\theta)}{H_{c2}^{\parallel}}\right)^2 = 1 \quad \text{(Eq.S2)}$$

To obtain an estimate of the in-plane coherence length, $\xi_{ab}$, the $H_{c2}$ vs T data were fit to the following models (*46*): The Gorter-Casimir (GC) two fluid model,

$$H_{c2}(T) = \frac{\Phi_o}{2\pi\xi_{ab}^2(T=0)}\left(1 - \left(\frac{T}{T_c}\right)^2\right) \quad \text{(Eq.S3)}$$

The Jones, Hulm, Chandrasekar (JHC) model combines the Ginzburg Landau formula for the upper critical field with the GC expression for the temperature dependence of the London penetration depth (*46*).

$$H_{c2}(T) = \frac{\Phi_o}{2\pi\xi_{ab}^2(T=0)} \frac{\left(1-\left(\frac{T}{T_c}\right)^2\right)}{\left(1+\left(\frac{T}{T_c}\right)^2\right)} \quad \text{(Eq.S4)}$$

and the Werthamer–Helfand–Hohenberg (WHH) model (*41*):

$$\ln(t) = \left(\frac{1}{2} + \frac{i\lambda_{SO}}{4\gamma}\right)\psi\left(\frac{1}{2} + \frac{\bar{h}+\frac{1}{2}\lambda_{SO}+i\gamma}{2t}\right) +$$

$$\left(\frac{1}{2} - \frac{i\lambda_{SO}}{4\gamma}\right)\psi\left(\frac{1}{2} + \frac{\bar{h}+\frac{1}{2}\lambda_{SO}-i\gamma}{2t}\right) - \psi\left(\frac{1}{2}\right) \quad \text{(Eq.S5)}$$

where:

$$t = \frac{T_{c0}}{T_c}, \quad \bar{h} = 2e\, H(v_F^2 \tau/6\pi T_c), \quad \gamma = \left( (\alpha\bar{h})^2 - \left(\tfrac{1}{2}\lambda_{SO}\right)^2 \right)^{\frac{1}{2}}, \quad \alpha = 3\hbar/2mv_F^2\tau$$

$$\lambda_{SO} = \hbar/3\pi T_c \tau_{SO}$$

τ is the transport scattering time and $\tau_{SO}$ is the spin-orbit scattering time. $H$ is the applied field, $v_F$ is the Fermi velocity. Fits to the data are given in Fig. S4. For the 4.4nm thick sample, the data is best fit with the GC model while the remaining films are best fit with the JHC model. WHH gives a poor fit in all cases.

## Estimation of London Penetration Depth

We obtain estimates for the London depth λ using the collective creep model (42,43) given by

$$U = U_o \ln(H/H_o) \quad (Eq.\,S6)$$

$$U_o = \phi_o^2 t/(256\pi^3 \lambda^2)$$

$$H_o \approx H_{c2}$$

Where $\phi_o$ is the flux quantum, $t$ is the sample thickness. The activation energy $U$ is determined from the slope of the linear region when plotting the resistive transitions on an Ahhrenius-type plot. We show an example of such analysis in Fig. S5**a** for a fixed parallel field of 0.6T with the transitions displayed taking place in various applied perpendicular fields. Once the activation energies are determined, they are plotted against the perpendicular field, and the London depth can be obtained from a linear fit using Eq. S6. We plot the full set of data for various fixed parallel fields in Fig. S5**b**.

## Normal State Transport

**Table S2. Summary of Transport parameters for samples from Series 1**

| $\upsilon_S/T_{c0}$ | 0 | $0.05T_c$ | $0.61T_c$ | $0.89T_c$ |
|---|---|---|---|---|
| $R_s$ (ohms/square) | 590 | 476 | 450 | 160 |
| $n_e$ (1/cm³) | 1.4e20 | 2.1e20 | 3.0e20 | 4.3e20 |
| $n_h$ (1/cm³) | 3.5e20 | 3.8e20 | 4.3e20 | 1.1e21 |
| $\mu_e$ (cm²/Vs) | 65 | 61 | 44 | 88 |
| $\mu_h$ (cm$^2$/Vs) | 42 | 44 | 40 | 58 |

Transport measurements were performed at 1.7K on the samples showing field-enhanced superconductivity in a Quantum Design Dynacool system. The variability in sample quality leads to differing values of the extracted charge densities and mobilities. These values are used to estimate the transport scattering times and constrain the fits of the superconducting data. As shown in our previous work on this material (22), the band structure consists of multiple crossings at the Fermi level and therefore to estimate the carrier density, we fit the data to the two carrier Drude model.

$$\rho_{xx} = \frac{1}{e} \frac{(n_h\mu_h + n_e\mu_e) + (n_h\mu_e + n_e\mu_h)\mu_h\mu_e H^2}{(n_h\mu_e + n_e\mu_h)^2 + (n_h - n_e)^2 \mu_h^2 \mu_e^2 H^2} \quad \text{(Eq.S7)}$$

$$\rho_{xy} = \frac{H}{e} \frac{(n_h\mu_h^2 - n_e\mu_e^2) + (n_h - n_e)\mu_h^2\mu_e^2 H^2}{(n_h\mu_e + n_e\mu_h)^2 + (n_h - n_e)^2 \mu_h^2 \mu_e^2 H^2} \quad \text{(Eq.S8)}$$

Values from the two carrier fits are given in Table S2. and the fits to the data for are given in Fig. S6. Even though this data was taken at 1.7K, a downturn can be seen at low field data due to incipient superconductivity. We therefore restricted the two carrier fit to the high field regime. The KF theory uses a free electron Fermi surface to approximate the electronic structure. If we use the above transport parameters with the Drude model with $m_e = 1$ to estimate the values of $p_F t \sim$ (5nm$^{-1}$ * 4.4nm) $\sim 20$ where $p_F$ is the Fermi momentum and $t$ is the sample thickness and the transport scattering time $\tau \sim 10^{-14}$ s $\sim 1000$K, we find they are comparable to the values from the fitting of the KF model (eee next section).

**Fitting of Critical Field Data to the Kharitonov-Feigelman Theory:**

In their original work, Abrikosov and Gorkov (AG) obtained (23) a formula for $T_c$ in the presence of generic pair-breaking perturbations of the form:

$$ln\left(\frac{T_{c0}}{T_c}\right) = \psi\left(\frac{1}{2} + \frac{\hbar v_s}{2\pi k_B T_c}\right) - \psi\left(\frac{1}{2}\right) \quad \text{(Eq.S9)}$$

where $v_s$ is the pair-breaking scattering rate. To take into account the polarization of magnetic impurity spins by the external magnetic field as well as the effects of spin-orbit scattering, disorder scattering, and the orbital (OE) and paramagnetic (PE) effects on $T_c$, Kharitonov and Feigelman (KF) (32) obtained a modified AG expression and with the following formula.

$$ln\left(\frac{T_{c0}}{T_c}\right) = \pi T \sum_\varepsilon \left(\frac{1}{|\varepsilon|} - C_0(\varepsilon)\right) \quad \text{(Eq.S10)}$$

Here, $C_0(\varepsilon)$ is the Cooperon Green function which satisfies:

$$\left(|\varepsilon| + v_S - \delta\Gamma(\varepsilon) + \frac{1}{2}(\hat{L}_o - \Gamma_{sf}(\varepsilon)) + \frac{3(h - n_S J \langle S_z \rangle)^2}{2v_{SOS}} + \frac{2(p_F t)^2 h^2}{9v}\right) C_0(\varepsilon) = 1 \text{ (Eq.S11)}$$

Where $v_S$ is the electron scattering rate of AG on magnetic impurities given by Eq. 2 of the main text, $\varepsilon = 2\pi T(n + \frac{1}{2})$, the fermionic Matsubara frequency, $\Gamma(\varepsilon) = v_S - \delta\Gamma(\varepsilon)$ is the full temperature-dependent exchange scattering rate and $\delta\Gamma(\varepsilon)$ gives reduction in the rate of scattering due to polarization of impurities. It is the enhancement of this term by the parallel field that leads to a reduction in the overall pair breaking rate. $\Gamma_{sf}(\varepsilon)$ is the rate of scattering *with* spin-flip, and

$$\hat{L}_o C_0(\varepsilon) = v_S \frac{\langle S_z \rangle}{S(S+1)} T \sum_\omega \frac{2\omega_S}{\omega^2 + \omega_S^2} C_0(\varepsilon - \omega)$$ 

where $\omega = 2\pi n T$ is the bosonic Matsubara frequency, and $\omega_S = gh$ where $g = 2$ and $h$ is the applied parallel field. See (32) for the full details of the above expressions. Matsubara sums were performed from $n \in [-500, 500]$. It was found that summing over a larger number of Matsubara frequencies did not change the fits. $v_{SOS}$ is the spin-orbit scattering rate resulting in changes in the spin-direction while preserving time-reversal symmetry. Note that this term in the Cooperon Green function is identical to that used in the KLB theory (*39*) with the notable change being the modified magnetic field which takes into account the exchange field from the magnetic impurities ($n_S J_{ex} \langle J_z \rangle$) as well as the externally applied field. In this expression, $J_{ex}$ is the exchange coupling constant. The sign of this exchange term determines whether the field generated by the polarized impurities enhances (ferromagnetic exchange) or opposes (antiferromagnetic) the applied field. $\langle J_z \rangle$ is the expectation value of the magnetic impurity spin in the z direction, and is given by the standard Brillioun function for paramagnetic impurities. $n_S$ is the concentration of magnetic impurities. The final term, which captures the orbital suppression of $T_c$, involves the transport scattering rate $v$ and the product of the sample thickness $t$ and Fermi momentum $p_F$, derived by considering diffusive transport of electrons.

In addition to the bare AG term, the various terms in Eq.S11 give rise to a renormalized electron scattering rate and therefore a modified Cooper pair lifetime and ultimately a modified $T_c$. Intuitively, terms that contribute negatively to the scattering rate tend to enhance $T_c$ while those contributing positively lead to a suppression. Full details of the various terms can be found in the original KF paper (*32*). We note that in the limit $h \to 0$, one recovers the original AG expression. Thus, one can interpret all other terms in the Cooperon expression as either augmenting or diminishing the pair-breaking scattering rate of the magnetic impurities. We obtained $T_c$ for a given field by numerically solving Eq.S10. Because the expression for the Cooperon involves a convolution term ($\hat{L}_o C_0(\varepsilon)$) that depends on knowledge of the Cooperon at all energies, we obtained the values of a Cooperon at a given ε self consistently. A trial Coooperon was given by AG formula $C_0(\varepsilon) = 1/(|\varepsilon| + v_S)$ *i.e.* the Cooperon with only unpolarized magnetic impurity scattering considered. This trial Cooperon was then used to compute the convolution term and give full Cooperon in the presence of all other effects (orbital, paramagnetic, polarization) by numerically solving (Eq.S11). The computed Cooperon was then used as the trial Cooperon in the next self-consistent step. This process was repeated until convergence between the trial Cooperon and the calculated Cooperon was reached. The error was obtained by computing the root-mean-square deviation between calculated and trial values. We used an error bound of $10^{-8}$ as it was found that the results did not change with tighter bounds. Convergence was reached within 5 self-consistent steps. Once the Cooperon for a given temperature was calculated, it was used in Eq.S10 to numerically determine the transition temperature. We verified the accuracy of the code by reproducing the calculations in (*32*).

According to the KF model, $p_F t$, $v$, $v_{sos}$, $v_s$ as well as an additional parameter $\varsigma = n_S |J| S / v_S$ are the parameters to fit the data, with the latter serving to determine the magnitude of the exchange field in the paramagnetic term. We can, however, constrain the values these parameters can take.

First, we fix $v_s$ based on the value of $T_c$ at zero field. Knowing $v_s$ allows us to estimate the magnitude of $J$ using eqn. 2 in the main text. The values for charge density from normal state transport allow us to estimate $N_F$ based on the Drude approximation along with approximating the Fermi surface to be that of the isotropic electron gas. The magnetic impurity concentration is taken to be that given in Fig. 2a. We find $J \sim 10$meV which translates to a value for $\varsigma \sim 1$. Per the derivation of KF, for Born scattering to hold, $\varsigma$ should be $\gg 1$. Our finding implies strong magnetic impurity scattering and the need to go beyond the Born approximation to fully capture the physics.

In the KF formula, $p_F t$ and $v$ are not independent. We fix $p_F t$ based on the results of the transport measurements and estimation of sample thickness and charge density, assuming an isotropic Fermi surface.

Because both the paramagnetic term as well as the orbital term contribute quadratically in $h$ to the pairbreaking, the parameters $v$ and $v_{sos}$, were highly correlated with one another. Similarly, the $\varsigma$ and $v_{sos}$ terms appear together in the same term in numerator and denominator, respectively. This meant that obtaining a unique fit to the data was challenging. However, for the parameters given in Table S3-S4, the data can be well reproduced using the model. The values for are $v$ and $v_{sos}$ are physically reasonable and are in agreement with what was found from the normal state transport.

Once the $0.89T_c$ sample was fit, $\varsigma$ was held fixed when fitting the other samples because this parameter is not expected to depend sensitively on sample quality. Using this value, we fit the remaining samples and obtain values for $v$, $v_{sos}$ which vary between samples but are within a factor of 3 of one another and can be explained by variations in sample quality (See Table S3,S4). We also find the expected relationship $v > v_{sos} > v_s$ to hold for the fit parameters.

Fits to a second series of samples is provided in Fig. S7 and obtained values of fit parameters are consistent between the two data sets.

Finally, a set of raw curves from the datasets showing field-induced superconductivity is given in Fig. S8.

**Table S3. Summary of fitting parameters Series 1**

| $c$ (at. %) | Undoped | 0.0075 | 0.014 | 0.019 |
|---|---|---|---|---|
| $\upsilon_S/T_{co}$ | 0 | 0.08 | 0.60 | 0.89 |
| $\upsilon/T_{co}$ | 1841 | 1534 | 1813 | 1032 |
| $\upsilon_{SO}/T_{co}$ | 99.0 | 139 | 174 | 348 |
| $\varsigma$ | 0 | 1.25 | 1.25 | 1.25 |
| $p_F t$ | 20 | 20 | 20 | 20 |

**Table S4. Summary of fitting parameters Series 2**

| $\upsilon_S/T_{co}$ | 0 | 0.56 | 0.75 | 0.84 | 0.89 |
|---|---|---|---|---|---|
| $\upsilon/T_{co}$ | 1641 | 1162 | 1311 | 1168 | 1394 |
| $\upsilon_{SO}/T_{co}$ | 67.3 | 59 | 59 | 143 | 178 |
| $\varsigma$ | 0 | 1.25 | 1.25 | 1.25 | 1.25 |
| $p_F t$ | 20 | 20 | 20 | 20 | 20 |

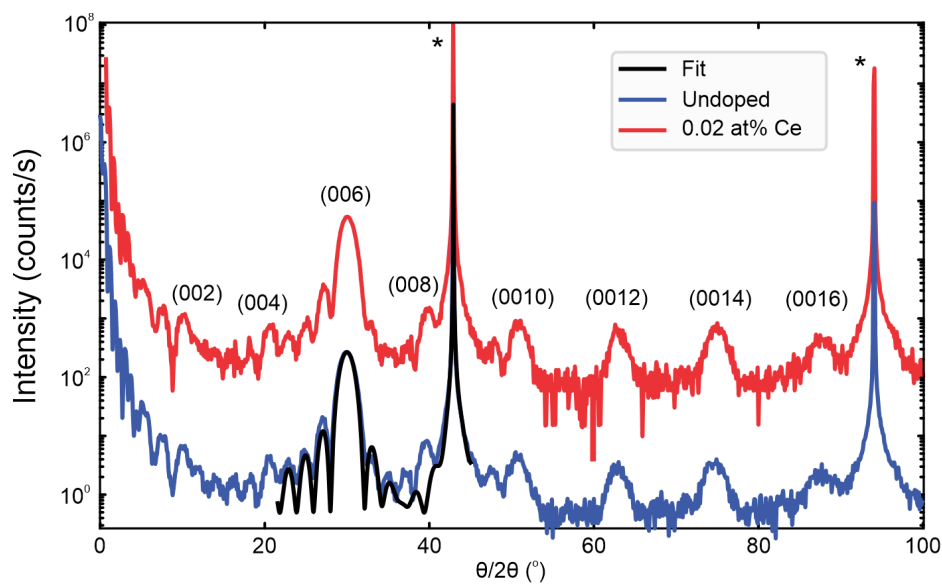

**Fig S1:** XRD comparison of undoped and 0.02 at% Ce doped ($0.89T_c$). Peaks are indexed according to the monoclinic $LaSb_2$ crystal structure. * indicates MgO {002} substrate peaks. Black line is a fit of the Laue fringes to estimate sample thickness. Fit yields $t$ = 4.4nm for both samples.

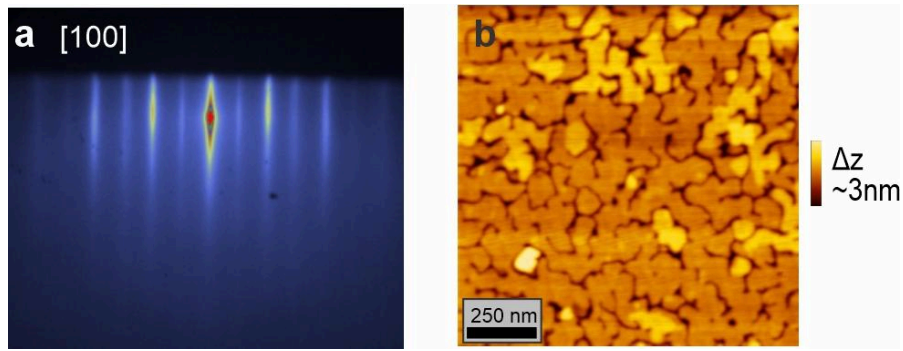

**Fig S2: a** Reflected high energy electron diffraction along the [100] direction of a LaSb$_2$ film with $t$~4.4 nm. **b** Atomic force microscopy of the surface of a $t$~4.4 nm film.

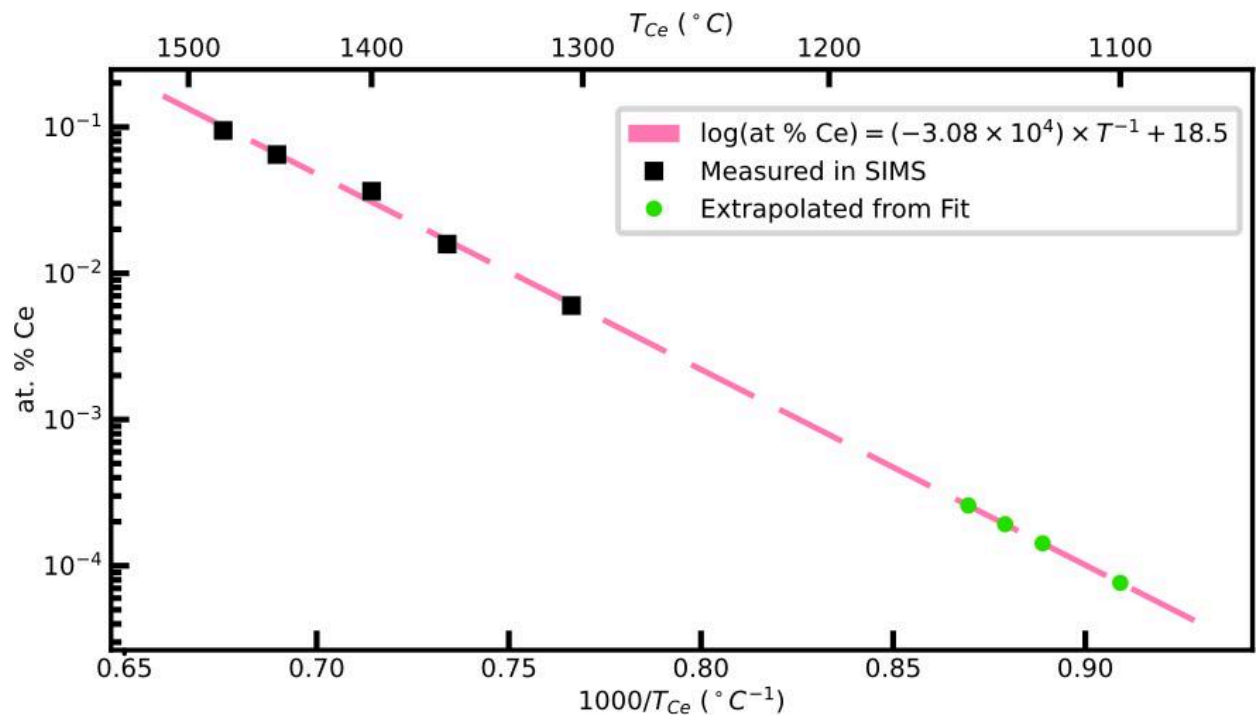

**Fig S3:** Secondary ion mass spectroscopy analysis used to estimate the concentration of cerium in dilute films.

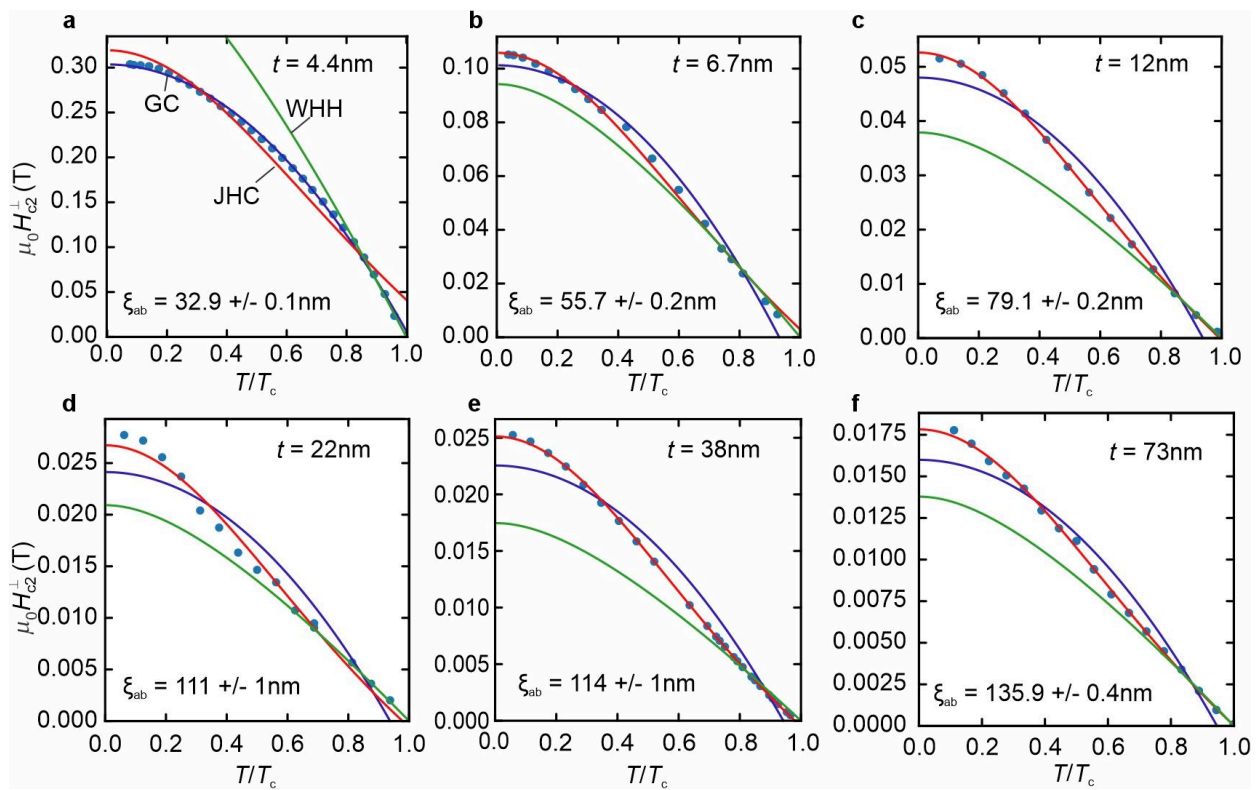

**Fig S4:** Fits of thickness dependent critical field data to various models

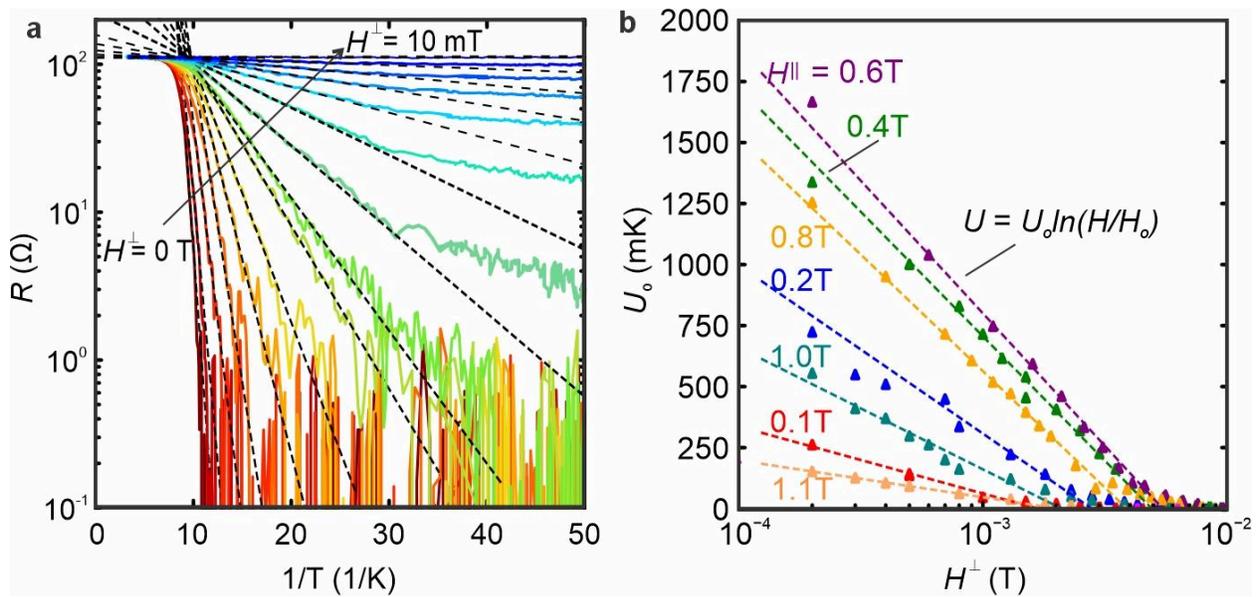

**Fig S5: a** Arrhenius plot of resistive transitions for $H_{\parallel}$ = 0.6T at various values of applied perpendicular field. Linear fits of the transitions are used to estimate activation energies **b** Extracted activation energies as a function of the perpendicular field for fixed applied parallel field.

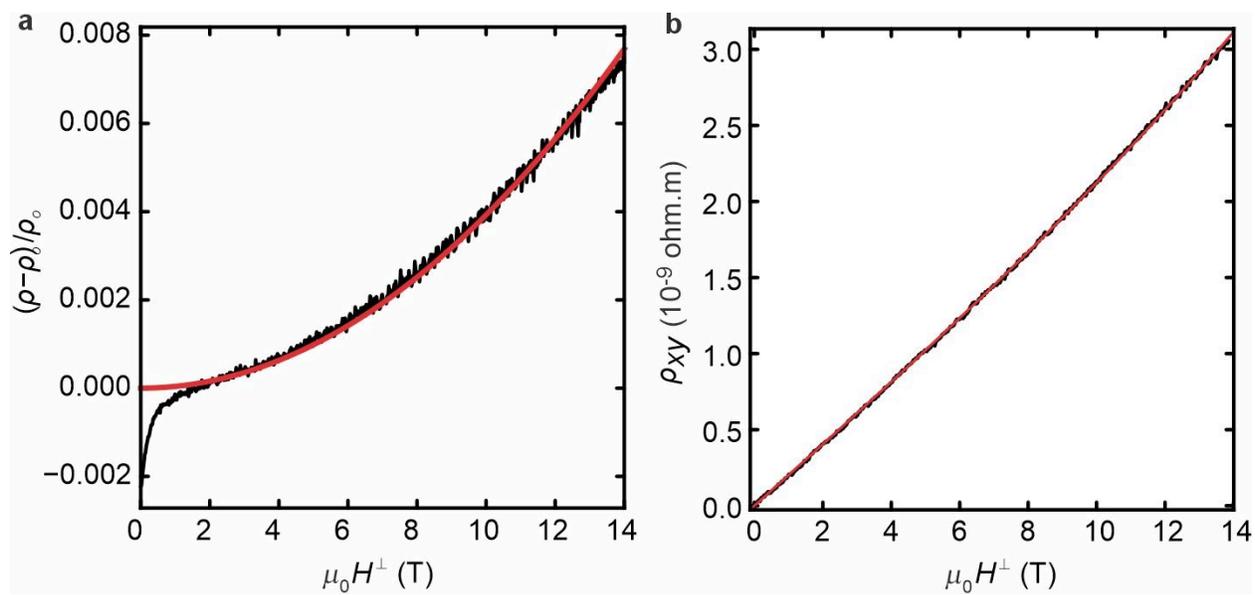

**Fig S6:** Two carrier fitting fitting to the $0.89T_c$ sample, taken at 1.8K

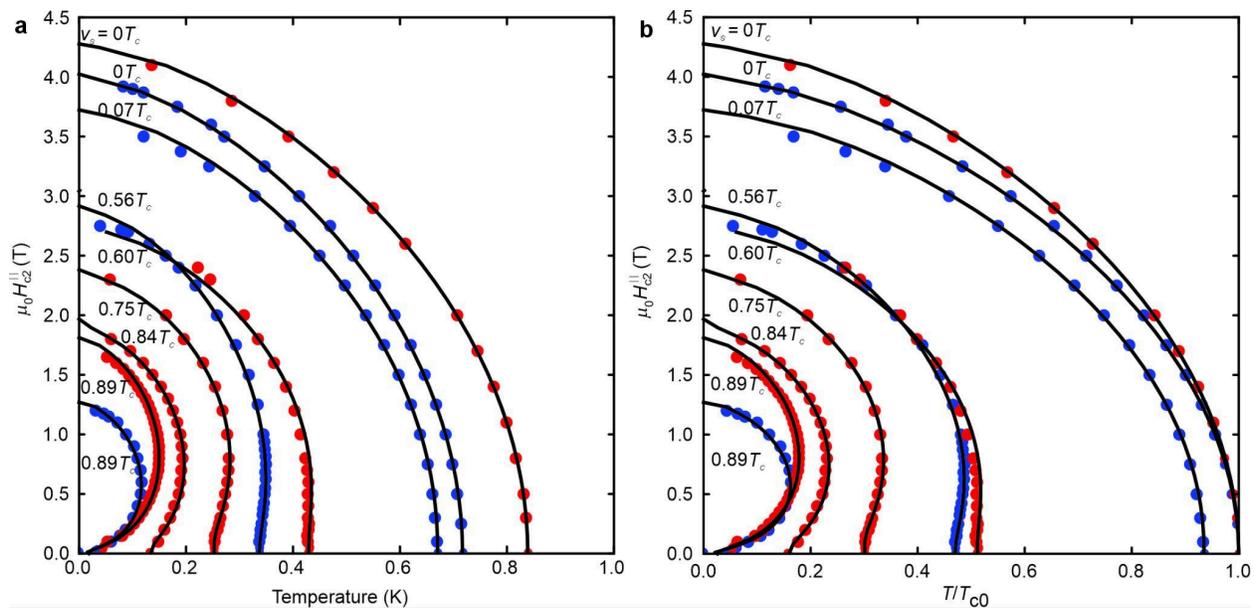

**Fig S7:** Fits to both sample series (Series 1 in blue, Series 2 in red), plotted in **a** as a function of $T$ and **b** $T/T_{c0}$.

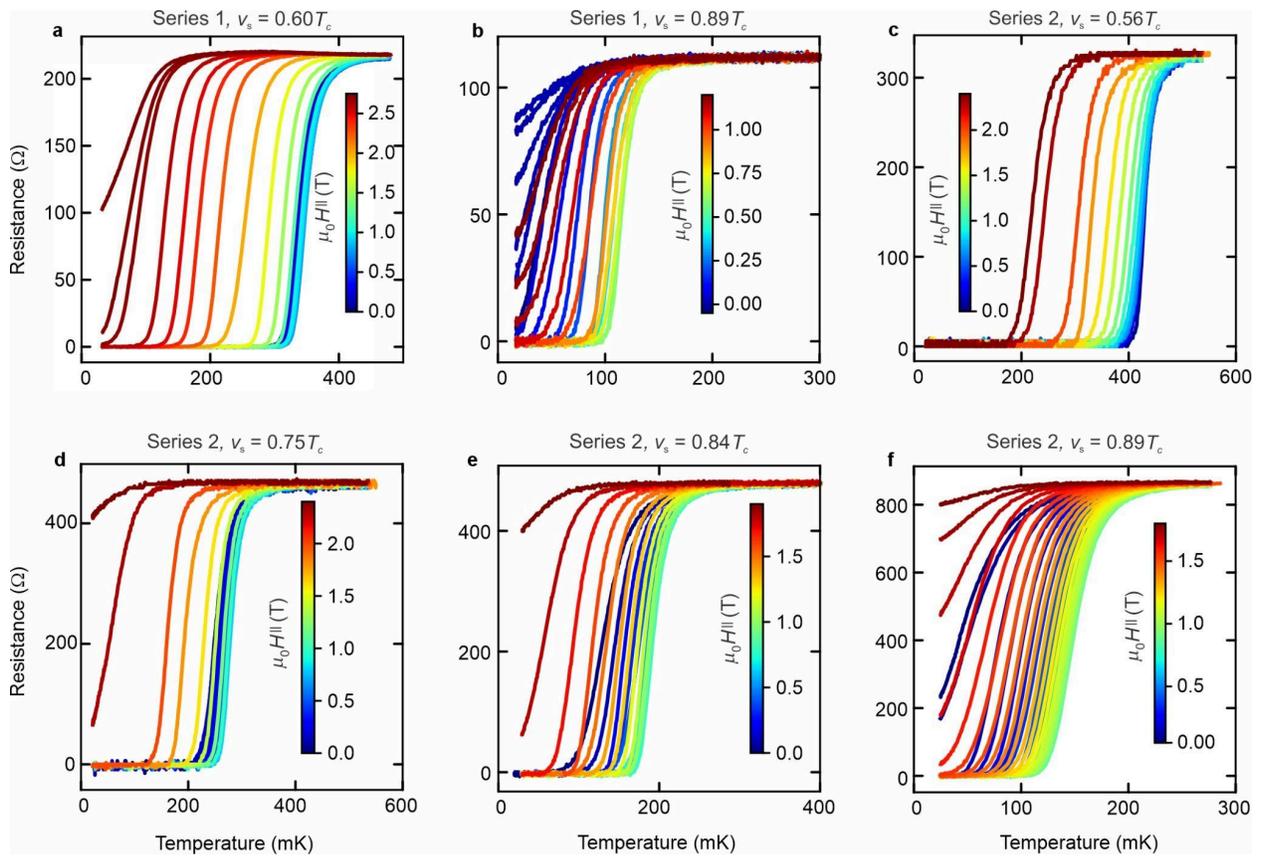

**Fig S8:** Additional raw data showing field-enhanced superconductivity in other samples from series 1 and 2.